\documentclass[twocolumn,prl,superscriptaddress,showpacs,preprintnumbers,amsmath,amssymb]{revtex4}

\usepackage{graphicx}
\usepackage{dcolumn}
\usepackage{bm}

\begin{document}
\title{Numerical and Theoretical Study of a Monodisperse
Hard-Sphere
Glass Former}
\author{P.~Charbonneau}\affiliation{Department of Chemistry,
Duke University, Durham, North Carolina, 27708, USA}
\author{A. Ikeda}\affiliation{Institute of Physics, University
of Tsukuba, Tennodai 1-1-1, Tsukuba 305-8571, Japan}
\author{J.~A.~van~Meel}\affiliation{FOM Institute for Atomic
and Molecular Physics, Science Park 113, 1098 XG Amsterdam, The
Netherlands}
\author{K. Miyazaki}\affiliation{Institute of Physics,
University of Tsukuba, Tennodai 1-1-1, Tsukuba 305-8571,
Japan}

\date{\today}
\begin{abstract}
There exists a variety of theories of the glass transition and
many more numerical models. But because the models need
built-in complexity to prevent crystallization, comparisons
with theory can be difficult. We study the dynamics of a
deeply supersaturated \emph{monodisperse} four-dimensional (4D) hard-sphere fluid, which has no such complexity, but whose
strong intrinsic geometrical frustration inhibits
crystallization, even when deeply supersaturated. As an
application, we compare its behavior to the mode-coupling theory (MCT) of glass
formation. We find MCT to describe this
system better than any other structural glass formers in lower
dimensions. The reduction in dynamical heterogeneity in 4D
suggested by a milder violation of the Stokes-Einstein relation
could explain the agreement. These results are consistent
with a mean-field scenario of the glass transition.
\end{abstract}
\pacs{64.70.qj, 61.43.Fs,64.70.pm, 66.30.hh} \maketitle

Though the conversation started generations ago, scientists
still debate the nature of the glass
transition~\cite{binder:2005}. The multiplicity of competing
frameworks even led some to declare that glass theories are
more numerous than glass theorists~\cite{chang:2008}. To be
fair, a great deal of progress toward a microscopic description
has been made over the last couple of decades, but many
fundamental questions remain unanswered. The intrinsic
complexity of many glass forming systems hinders progress. To avoid interference from precipitous crystallization,
simulated fluids require built-in complexity:
bidisperse~\cite{bernu1985,Foffi2004}, polydisperse~\cite{Kumar2006} or
multicomponent~\cite{kob:1994} mixtures,
anisotropic~\cite{Shintani2006} or
frustrated~\cite{dzugutov:2002} pair interactions, etc. The
situation is even more intricate in experiments, where good
glass formers are, with only one reported
exception~\cite{bhat:2007}, molecular liquids, polymer melts,
or metallic alloys. Quantitative comparisons between the
microscopic theories and these systems are formidably
challenging, which impedes assessing the validity and
limitations of the predictions~\cite{Cavagna2009}. A plain
glass former, {\it i.e.}, a monatomic, one-component liquid with an interaction potential as simple as possible would thus be
greatly beneficial for the field's progress.

\begin{figure}
\center{\includegraphics[width=1.03\columnwidth]{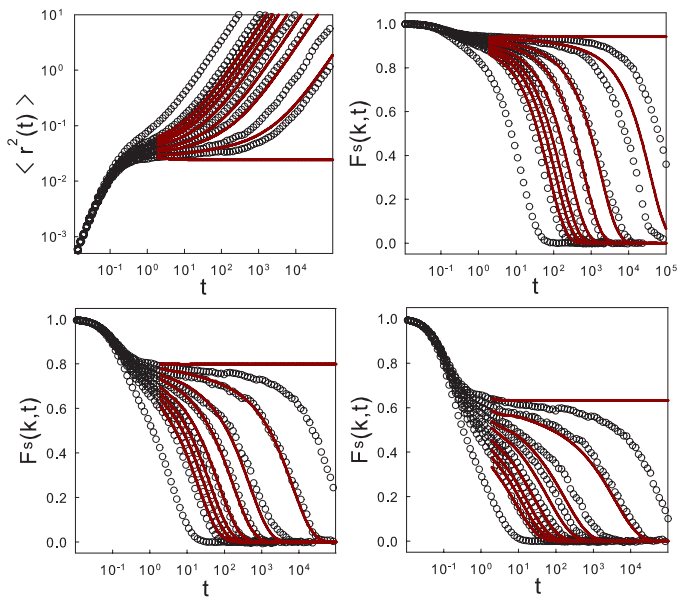}}
\caption[$\langle r^2(t)\rangle$ and $F_s(k,t)$]{(Color online)
(a) MSD and (b) $F_s(k,t)$ at $\eta=0.370$, $0.386$, $0.389$,
$0.392$, $0.395$, $0.398$, $0.401$, $0.404$, and $0.407$, from
left to right. Solid lines are MCT fits using $\varepsilon$ as
control parameter (see text). At the highest density
$\tau_\alpha$ deviates from the power law scaling of
Fig.~\ref{fig:gamma}. } \label{fig:msd}
\end{figure}

In this Letter, we report numerical simulations of a glass
former made of monodisperse 4D hard spheres (HS). Monodisperse HS have a single non-trivial thermodynamic parameter, the
volume fraction $\eta$, which makes them the simplest model of
isotropic fluids and crystals. 
Surprisingly, unlike in 2D and 3D, monodisperse HS in 4D
and higher dimensions are strongly geometrically frustrated
with respect to the crystal~\cite{vanmeel:2009b}. The
simplex-based liquid structure is geometrically distinct from
the crystalline order, which inhibits nucleation and
facilitates glass formation~\cite{skoge:2006,vanmeel:2009}.
Compression studies of 4D HS suggest that the
structural relaxation timescale becomes longer than the slowest
accessible quenching rate around
$\eta\approx0.40-0.41$~\cite{skoge:2006,Parisi2008b}. A rough
estimate gives the nucleation rate of reasonable system sizes
for simulations to be eight to ten orders of magnitude slower
than the structural relaxation
timescale~\cite{vanmeel:2009,Auer2001}. In monodisperse 3D
HS, in contrast, at $\eta\approx0.54$ multiple crystal
nuclei form simultaneously on the structural relaxation
timescale, while the onset of slow dynamics is generally agreed to be $\eta\approx 0.58$. Bidisperse or polydisperse HS are thus used to study glass formation
(e.g.~\cite{Foffi2004,Kumar2006}). We perform molecular
dynamics simulations with the event-driven package
of Ref.~\cite{skoge:2006} in a system of $N=4096$ particles, in
order to examine the 4D system's glass forming
properties~\cite{footnote:1}. We expect finite-size effects to
be small in this regime, as is the case at similar
supersaturations in 3D for $N=512$~\cite{Karmakar2009}. 
The structure factor
$S(k)$ of the deeply supersaturated fluid remains liquid like
at all densities, but the complete absence of crystallinity is
also checked by an order parameter developed to detect
nucleation~\cite{vanmeel:2009}. Dynamically, the two-step
growth of the mean-square displacement (MSD)
$\langle |{\bf r}_i(t)-{\bf r}_i(0)|^2 \rangle$, where $\mathbf{r}_i(t)$ is the position of the $i$-th
particle, shows a lengthening caging plateau with density, a signature of structural glass
formers (Fig.~\ref{fig:msd}).

As an application of this simple model glass former, we
consider the role of dimensionality in glass formation, which
is a subject of considerable theoretical discussion~\cite{Biroli2007b,Parisi2008b,Kirkpatrick1987}. The abrupt dynamical slowdown near the glass transition suggests the
presence of a kinetic and/or a hidden thermodynamic
singularity. The most direct evidence for such a singularity is
the growth of spatio-temporal fluctuations on the structural
relaxation timescale, which results in fast and slow moving
regions in supercooled liquids. This dynamical heterogeneity is
typically monitored through four-point correlation
functions~\cite{Toninelli2005b} and Stokes-Einstein (SE)
relationship violations~\cite{binder:2005,Cicerone1996}. As for conventional
continuous phase transitions, the impact of these fluctuations
should be reduced in higher dimensions as the system becomes
more mean-field-like. Comparing systems of different
dimensionality should allow to better understand the glass
transition as a critical phenomenon and to test this mean-field scenario. A first attempt in this direction
was recently made by Eaves and Reichman for a 4D binary
Lennard-Jones (BLJ) model system~\cite{eaves:2009}, but the
complication of identifying the dimensional correspondence
between BLJ systems makes quantitative comparisons difficult.


Insights into the mean-field scenario of the glass transition
are gained by a MCT analysis of the dynamical results.
A mean-field description of the thermodynamic ``ideal'' glass
transition based on the replica theory suggests that HS undergo a dynamical transition at $\eta_c$, before
reaching the
thermodynamic
glass transition of the one-step replica
symmetry breaking at
$\eta_K>\eta_c$~\cite{mezard1999,Parisi2008b}.
This decoupling between dynamical and thermodynamic anomalies is
demonstrated by simulation~\cite{Santen2000}.
MCT is conjectured to be the
dynamical counterpart of this approach below $\eta_c$, because
its mathematical structure is equivalent to the dynamical
equations of a mean-field $p$-spin glass
model~\cite{Kirkpatrick1987d} for which the relation between
the dynamical and the thermodynamic glass transitions is
rigorously established~\cite{Castellani2005}. Though
contentious, MCT is one of the most successful theories of the
glass transition. It uses static structural information, such
as the radial distribution function $g(r)$, to provide
first-principles predictions of the slow dynamics of fluids
before their dynamical arrest~\cite{gotze:2009}. At mild
supercooling it qualitatively captures the onset of the
two-step decay of time correlation functions and the algebraic
relaxation of the intermediate time regime. On approaching
$\eta_c$ it predicts a power-law divergence of the structural
relaxation time instead of the well-known Vogel-Fulcher-Tammann
behavior~\cite{binder:2005}. This power law describes
simulation and experimental observations fairly well for a
range of densities below the $\eta_c$ fitted from the dynamical
data~\cite{gotze:2009}, but the divergence at $\eta_c$ is
rounded off due to activated events, which the theory does not
capture. A failing of MCT is that it foretells a nonergodic
freezing of the dynamics at a much lower $\eta_c$ (or higher
temperature $T_c$ for thermal systems) than the experimental
and simulation glass transition point $\eta_g$ (or
$T_g$)~\cite{binder:2005}. 
Moreover the nonergodic freezing point 
obtained by fitting the simulation data with the MCT power law are systematically lower than what the theoretical predicts~\cite{Kob2002,Brumer2004}. MCT also lacks an
explanation for the violation of the SE relation, which is
mostly attributed to strong dynamical heterogeneity near
the glass transition and is missing in the
theory~\cite{Mayer2006b}. Yet if MCT is indeed a dynamical
mean-field theory, a reduction in the degree of heterogeneous
dynamics by increasing dimensionality should improve the
its agreement with simulation results.

MCT is expressed as a series of nonlinear integro-differential
equations for correlation functions such as the intermediate
scattering function $F(k,t)=N^{-1}\langle \delta\rho_{\bf k}(t)
\delta\rho_{\bf k}^{\ast}(0)\rangle$, where $\delta\rho_{\bf
k}(t)$ is the density fluctuation in reciprocal space.
Generalization of the theory to 4D straightforwardly
gives~\cite{Bayer2007b}
\begin{equation}
\ddot{F}(k,t) + \Omega_k^2 F(k,t) + \int^t_0\!\!\!\! ds \
M(k,t\!-\!s) \dot{F}(k,s) = 0, \label{eq:MCT}
\end{equation}
where $\Omega_k^2\equiv k_BT k^2/mS(k)$, $S(k)\equiv F(k,t=0)$
is the static structure factor, and $M(k,t)$ is the memory
kernel. $M(k,t)$ can be further decomposed into fast and slow
components $M(k,t)=$ $M_{\mbox{\scriptsize fast}}(k,t)$ $+$
$M_{\mbox{\scriptsize MCT}}(k,t)$ with
\begin{equation}
\begin{aligned}
& M_{\mbox{\scriptsize MCT}}(k,t)= \int_{0}^{\infty}\!\!\!\!
\mbox{d}q \int^{|q+k|}_{|q-k|}\!\!\!\!\!\!\!\!\mbox{d}p~
V_k(q,p) F(q,t)F(p,t),
\end{aligned}
\label{eq:MCT.memory}
\end{equation}
where $V_k(q,p)$ $\equiv$ ${\eta}$ $\sqrt{4k^2p^2 -f_{+}^2}$
$[f_{+}c(q)+f_{-}c(p)]^2$ $/$${16\pi k^4}$, $c(k)\equiv
\{1-1/S(k)\}/\rho$ is the direct correlation function,
$\rho=N/V$ is the number density, and $f_{\pm} \equiv k^2 \pm
(q^2 - p^2)$. Binary collisions dominate the fast decaying part of the memory kernel $M_{\mbox{\scriptsize fast}}(k,t)$, which is conventionally determined by fitting
simulation data~\cite{Kob2002}. Here, the analysis is done for
the self part of the intermediate scattering function
$F_s(k,t)\equiv \left\langle e^{i
\mathbf{k}\cdot[\mathbf{r}_i(t)-\mathbf{r}_i(0)]}\right\rangle
$ for which the MCT expression is similar to
Eq.~\ref{eq:MCT.memory}~\cite{Bayer2007b}.
Because the MCT analysis is very sensitive to the details of the $S(k)$ input,
particular care is taken to interpolate and extrapolate (to
larger $k$) the simulation data to a continuous functional
form. We extend the simulated $g(r)$ for distances larger than
half the simulation box with a damped oscillatory function. The
Fourier transform of $g(r)$ is in good agreement with the
direct computation of $S(k)$. We compute (see
Fig.~\ref{fig:msd}) $F_s(k,t)$ for several wavevectors,
one of which ($k=8.3$) is close to the first peak of the
structure factor and thus not too far from the dominant
microscopic caging and relaxation length
scales~\cite{charbonneau:2007}. The decorrelation of density
fluctuations displays the characteristic caging plateau and
allows for the extraction of the structural relaxation time
$\tau_{\alpha}$, defined as $F_s(k,\tau_{\alpha}) = 1/e$.

\begin{figure}
\includegraphics[width=0.9\columnwidth]{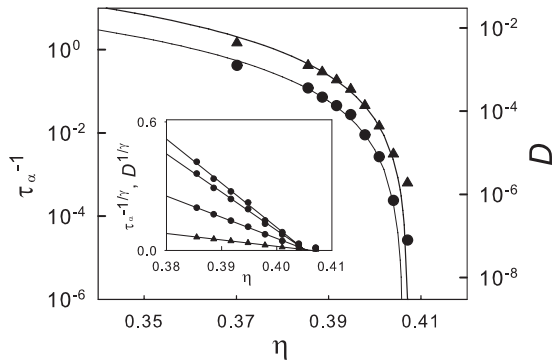}
\caption{Power-law fit of $D$ ($\blacktriangle$) and
$\tau_\alpha^{-1}$ for $k=8.3$ ($\bullet$). Inset:
$D^{1/\gamma}$ and $\tau_{\alpha}^{-1/\gamma}$ for $k=8.3$,
$7.5$, and $4.2$, from top to bottom, scale linearly with a
shared intersection point. } \label{fig:gamma}
\end{figure}

MCT predicts (Eq.~\ref{eq:MCT}) that both $\tau_\alpha$ and the
diffusion coefficient $D$ scale with the same power-law
exponent $\gamma$ and critical density $\eta_c$, {\it i.e.},
$\tau_\alpha^{-1}, D \propto  |\eta_c-\eta|^{\gamma}$, with
$\gamma^{\mbox{\scriptsize MCT}}=2.8$ and
$\eta_{c}^{\mbox{\scriptsize MCT}}=0.379$. In
Fig.~\ref{fig:gamma}, we fit the simulation data to this form
by fixing $\gamma^{\mbox{\scriptsize MCT}}$ and leaving
$\eta_c$ free. All the data except for the densest system
follow a power law with $\eta_{c, D}^{\mbox{\scriptsize
sim}}=0.407$ and $\eta_{c,\tau_\alpha}^{\mbox{\scriptsize
sim}}=0.406$ for all wavevectors. The discrepancy between
$\eta_{c, D}^{\mbox{\scriptsize sim}}$ and
$\eta_{c,\tau_\alpha}^{\mbox{\scriptsize sim}}$ is smaller
(0.2\%) than for 3D HS (0.5\%)~\cite{Foffi2004}, and
the concordance with $\eta_{c}^{\mbox{\scriptsize MCT}}$
slightly improves. It is interesting to note that
$\eta_c^{\mbox{\scriptsize sim}}$ is very close to the
extrapolated dynamical arrest point independently obtained at
very slow compression rates~\cite{skoge:2006,Parisi2008b}. For
consistency check, we also fit $\gamma$ while fixing
$\eta_{c}^{\mbox{\scriptsize sim}}=0.406$, which gives
$\gamma_{D}^{\mbox{\scriptsize sim}}=2.4$ and
$\gamma_{\tau_\alpha}^{\mbox{\scriptsize sim}}=2.7$ for all
wavevectors. In contrast, for 3D
BLJ~\cite{kob:1994,Flenner2005e}, which is the best
characterized system, $\gamma_{D}^{\mbox{\scriptsize
sim}}\approx 1.8$ and $\gamma_{\tau_\alpha}^{\mbox{\scriptsize
sim}}\approx 2.3$, whereas $\gamma^{\mbox{\scriptsize
MCT}}=2.46$. Another notable feature is that the power law fits
over almost four time decades, reaching densities where
$\varepsilon \equiv 1- \eta/\eta_c^{\mbox{\scriptsize sim}}$ is
less than $0.5 \%$ (Fig.~\ref{fig:gamma}). In 2D and 3D systems
the structural relaxation timescale follow a power law for only
two or at most three
decades~\cite{kob:1994,Flenner2005e,Kumar2006,foffi:2003,berthier:2009},
before the theoretical description breaks down because of
activated processes. 

We also consider how well MCT describes the full MSD and
$F_s(k,t)$ curves, using $\varepsilon$ as input parameter to
scale out the $\eta_c$ dependence (Fig.~\ref{fig:msd}). To
tease out the long-time MCT behavior from Eq.~\ref{eq:MCT}, the
short-time $F_s(k,t)$ decay (up to $t \approx 10^{0}$) is
imported from the simulation curves. This procedure is
equivalent to, but more direct than the standard fit of
$M_{\mbox{\scriptsize fast}}(k,t)$ from
simulation~\cite{Kob2002}. The concordance between simulation
and calculated $F_s(k,t)$ is almost perfect for $k=8.3$ and
$11.9$, up to densities where $\tau_\alpha$ deviates from the
power-law behavior and MCT foretells the nonergodic transition.
The calculated MSD also matches the simulated curves quite
well, except for $\eta=0.404$ and above.
The correspondence between simulation and MCT is noticeably better
than for lower-dimensional systems, where for every relaxation
curve both $\eta_c$ \emph{and} $k$ must be rescaled, in order
to achieve a reasonable collapse~\cite{Voigtmann2004}. Besides
the power-law scaling, MCT successfully describes several other
features in 3D, such as the time-temperature superposition, the
von-Schweidler law at the beta relaxation regimes, the
$k$-dependence of the plateau height (the non-ergodic
parameter), the exponents of stretched exponential relaxation
at the alpha relaxation regime $\beta(k)$, and the beta
relaxation exponent $b$~\cite{gotze:2009}. The concordance
between theory and simulation for all of these features is at
least as good for 4D HS as for 3D systems.

\begin{figure}
\includegraphics[width=0.9\columnwidth]{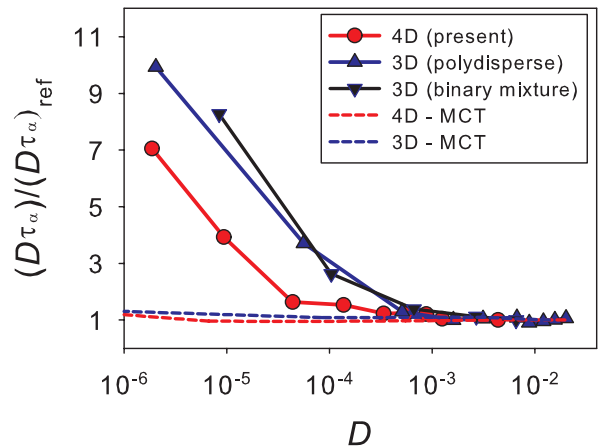}
\caption[SE Violation] { (Color online) Violation of the SE
relation in 4D monodisperse and 3D bidisperse~\cite{Foffi2004}
and polydisperse~\cite{Kumar2006} HS, where
$D\tau_\alpha$ is normalized by its low density
$(D\tau_\alpha)_{\mbox{\scriptsize ref}}$ value. MCT predicts
only a very small violation of the SE relation for both 3D and
4D systems.} \label{fig:tbd}
\end{figure}
The quantitative improvement of MCT predictions in 4D over 3D
suggests that fluctuations are less important in higher
dimensions. We analyze the system's dynamical heterogeneity
through the SE relation to verify this hypothesis.
Figure~\ref{fig:tbd} shows $D\tau_\alpha$ for the full range of
densities explored along with the corresponding 3D HS
results~\cite{Kumar2006,Foffi2004}. In order to put the
different dimensions on an equal footing, we plot
$D\tau_\alpha$ against the diffusion coefficient $D$. The SE
relation holds for diffusivities over a decade smaller in 4D
than in 3D (up to $\varepsilon \approx 1\%$),
which is slightly before $D$ and $\tau_\alpha$ start deviating
from the MCT-predicted power law in Fig.~\ref{fig:gamma}. If a
similar degree of polydispersity were used in 4D as in 3D we
expect the SE violation to be suppressed even
more~\cite{Kawasaki2007}. The suppression of the SE relation
violation and the better agreement of all available dynamical
observables with MCT predictions suggest that 4D HS
are dynamically more mean-field-like than the 2D and 3D
equivalents. The improvement of the agreement with MCT is
however incremental, which hints that if an upper critical
dimension $d_c$ for the glass transition exists, it is larger
than four. This result is consistent with general theoretical
arguments that give $d_c=8$~\cite{Biroli2007b,Biroli2006b}.
Biroli {\it et al}. also obtained that the SE relation
violation should scale as
$D\tau_\alpha\sim\varepsilon^{d/4-2}$~\cite{Biroli2007b}. The
growth in Fig.~\ref{fig:tbd} is {\it not} inconsistent with
this scaling, but the SE relation violation for the range of
densities we explore is too mild to be conclusive about the
exponent. A more direct measure of dynamical heterogeneity
would be to compute the four-point correlation function
$\chi_4(t)$, but the lack of statistical accuracy and the
relatively small system sizes prevent us from reporting the
results here. We will consider $\chi_4(t)$ in future studies.

The 4D monodisperse HS fluid we study is quite
convenient to examine the glass transition. Its simplicity and
slow nucleation rate allow high-accuracy comparisons of its
glass-forming properties with microscopic theories. The
agreement of the system with MCT, which is broader than for any
lower-dimensional equivalents, and the strong suppression of SE
relation violation are consistent with the dynamical mean-field
scenario of the glass transition~\cite{Parisi2008b}. The
results also suggest that 4D is still below the upper critical
dimension, if it exists, because the fluctuations due to
activated processes round off the sharp dynamical singularity.
Given that even the mean-field picture has not been rigorously
established for the structural glass transition, the study of
clean, higher-dimensional systems such as 4D HS is
likely to play a crucial role in assessing the validity and
limitations of the various glass theories.

\acknowledgments

We thank D. Reichman and F. Zamponi for stimulating
discussions. PC acknowledges computer time at the Dutch center
for high-performance computing SARA and startup funding from
Duke University. This work is partially supported by
Grant-in-Aid for JSPS Fellows (AI), KAKENHI; \# 21540416, (KM),
and Priority Areas ``Soft Matter Physics'' (KM).


\begin{thebibliography}{30}
\bibitem[{foo()}]{footnote:1} \bibinfo{note}{The particle
    diameter $\sigma$ sets the unit of length, the
  thermal energy $k_BT$ the energy unit, $m$ the unit of mass,
  and time $t$ is expressed in $(k_BT/m\sigma^2)^{1/2}$ units.
  At least eight independent initial configurations are
  equilibrated
after a slow compression from the ideal gas limit. }
\end{thebibliography}
\end{document}